\documentstyle[11pt]{article}

\def\be{\begin{equation}}  
\def\ee{\end{equation}} 
\def\bdm{\begin{displaymath}}
\def\edm{\end{displaymath}} 
\def\ket#1{| #1 \rangle}
\def\bra#1{\langle #1 |} 
\def\BraKet#1#2{\langle #1 | #2 \rangle}

\begin{document}
\centerline{\large\bf Relativistic quantum coin tossing} 
\vskip 3mm
\centerline{S.N.Molotkov and S.S.Nazin} 
\vskip 2mm
\centerline{\sl\small Institute of Solid State Physics of Russian Academy of Sciences,}
\centerline{\sl\small 142432 Chernogolovka, Moscow District, Russia\\
e-mail: molotkov@issp.ac.ru, nazin@issp.ac.ru} \vskip 3mm

\begin{abstract}
A relativistic quantum information exchange protocol is proposed
allowing two distant users to realize ``coin tossing'' procedure.
The protocol is based on the point that in relativistic quantum theory 
reliable distinguishing between the two orthogonal states generally requires
a finite time depending on the structure of these states.
\end{abstract}
PACS numbers: 03.67.-a, 03.65.Bz, 42.50Dv

The coin tossing protocol is one of the simplest cryptographic protocol
and can be described in the following way. Suppose that two mistrustful 
parties, A and B, wish to produce a random bit, e.g. employing a coin
tossing procedure or random numbers generator producing 0 or 1 with equal 
probabilities. Zero outcome means that the participant A won, and outcome
1 means that he lost. If A and B are not spatially separated 
the task is trivial. However, if A and B are located at distant sites
and can only exchange information through a communication channel the 
problem could even seem unsolvable since both A and B seem to be able to
cheat without being detected.

For the case when A and B can only exchange information through a classical
communication channel the problem was solved by Blum [1]. Strictly speaking,
the protocol suggested in Ref.[1] is not secure against the cheating of one of 
the parties since it is based on the unproven computational complexity
of the discrete logarithm problem [1]. For example, if one of the participants
had a quantum computer (which has not yet been actually built), he could
always win due to a fast computation of the discrete logarithm [2,3].

However, if there exists a quantum communication channel between users
A and B, it is possible to realize various information exchange protocols
whose security is based on the fundamental laws of nature (quantum theory) 
rather than the computational complexity. Different protocols have been 
suggested and studied so far: quantum key distribution [4--6], 
quantum bit commitment [7--9], quantum coin tossing [10], quantum 
gambling [11], and quantum secret sharing [12].

It was shown earlier that the ideal quantum coin tossing protocol
is impossible in the framework of the non-relativistic quantum mechanics [13,14].
(The protocol is said to be ideal if the probability of accepting 
of absence of cheating by both parties is exactly one, and both
outcomes 0 and 1 occur with equal probabilities of 1/2.)
However, a protocol can be designed in which the absence of cheating is accepted
by both parties with a probability arbitrarily close to one [10].

Recently, a bit commitment protocol and coin tossing protocol were proposed
which take into account the finite speed of signal propagation. 
In these protocols, the information is carried by the classical states.
These protocols assume that the two parties A and B each have a couple
of spatially separated sites fully controlled by them 
(for details, see Ref.[15]).
In our opinion, this scheme implicitly assumes the existence of a
communication channel between the sites $A_1$ and $A_2$ 
(as well as between $B_1$ and $B_2$) which is secure against the 
substitution of information transmitted through it (impersonation)
or reqires prior sharing of a key string (or afterwards physically 
getting together to compare notes).

Proposed below is an example of the real time relativistic 
coin tossing protocol.

All quantum cryptographic protocols actually employ the following two
features of quantum theory. The first one is the no cloning theorem [16], 
i.e. the impossibility
of copying of an arbitrary quantum state which is not known beforehand or,
in other words, the impossibility of the following process:
\bdm
\ket{A}\ket{\psi} \rightarrow U(\ket{A}\ket{\psi})=
                 \ket{B_{\psi}}\ket{\psi}\ket{\psi},
\edm
where $\ket{A}$ and $\ket{B_{\psi}}$ are the
apparatus states before and copying act, respectively, and
$U$ is a unitary operator. Such a process is prohibited
by the linearity and unitary nature of quantum evolution.
Actually, even a weaker process of obtaining any information about
one of the two non-orthogonal states without disturbing it is impossible,
i.e. the final states of the apparatus $\ket{A_{\psi_1}}$ 
and $\ket{A_{\psi_2}}$ corresponding to the initial input states
$\ket{\psi_1}$ and $\ket{\psi_2}$, respectively, after the unitary 
evolution $U$, 
\bdm
\ket{A}\ket{\psi_1} \rightarrow U(\ket{A}\ket{\psi_1})=
                         \ket{A_{\psi_1}}\ket{\psi_1} ,
\edm
\bdm
\ket{A}\ket{\psi_2} \rightarrow U(\ket{A}\ket{\psi_2})=
                         \ket{A_{\psi_2}}\ket{\psi_2} ,
\edm
can only be different, $\ket{A_{\psi_1}} \neq \ket{A_{\psi_2}}$, if 
$\BraKet{\psi_1}{\psi_2} \neq 0$ [17], which means the impossibility 
of reliable distinguishing between non-orthogonal states.
There is no such a restriction for orthogonal states.
That is why almost all cryptographic protocols employ
non-orthogonal states as information carriers, the only 
exception being the protocol suggested in Ref.[18].

{\it Two orthogonal states can be reliably distinguished, and within the
framework of non-relativistic quantum mechanics this can be done 
instantly.} It is this circumstance that is actually behind the 
impossibility of designing a cryptographic protocol based on a pair
of orthogonal states within the framework of non-relativistic 
quantum mechanics.

However, in the relativistic quantum field theory the situation is different.
The physical field observables associated with the two points separated by a 
space-like interval cannot have any causal relations and the commutator
of field operators is zero outside the light cone [19]:
\be 
[u^-(\hat{x}_1),u^+(\hat{x}_2)]_{\pm}=-iD^-(\hat{x}_1-\hat{x}_2), 
\ee
where $u^{\pm}(\hat{x})$ are the field operators, $\hat{x}_{1,2}$ are
the points of the four-dimensional space-time, and $D^-(\hat{x}_1-\hat{x}_2)$ 
is the negative-frequency commutator function [19]. This circumstance imposes 
a restriction on the time required for a reliable (in a single measurement act)
distinguishing of a pair of orthogonal states.

Before describing the protocol, we shall first discuss the states and 
measurements it employs. Any one-particle state of the field can be represented 
in the form
\be
\ket{\psi_{1,2}}=\int\psi_{1,2}(\hat{p})\delta(\hat{p}^2-m^2)u^+(\hat{p})d\hat{p}\ket{0},
\ee
where the integration is performed over the mass surface,
$\psi_{1,2}(\hat{p})$ is the field amplitude, and 
$\ket{0}$ is the vacuum state. In the rest of the paper we shall deal with
the massless particles (e.g. photons). Therefore, the
field operator $u^+(\hat{p})$ will be interpreted as the creation operator 
of a photon in the Coulomb gauge. We shall also assume that the amplitudes 
$\psi_{1,2}(\hat{p})$ are chosen in such a way that the states $\ket{\psi_{1,2}}$ 
are orthogonal:
\be
\BraKet{\psi_1}{\psi_2}=\int\int\psi^*_1({\bf p^{'}})\psi_2({\bf p})
\bra{0}u^-({\bf p^{'}})u^+({\bf p})\ket{0}
\frac{d{\bf p^{'}}d{\bf p} } { \sqrt{ 2p_{0}^{'} }\sqrt{ 2p_0} }=
\int\psi^*_1({\bf p})\psi_2({\bf p})\frac{d{\bf p}}{2p_0}=0,
\ee
\bdm
[u^-({\bf p'}),u^+({\bf p})]_-=\delta({\bf p'-p}).
\edm
In contrast to the non-relativistic quantum mechanics, a detailed consistent
theory of measurement in quantum field theory has not yet been developed.
For the one-particle states, we shall take advantage of the analogy with
the non-relativistic case. A measurement allowing to distinguish between
the two orthogonal states is given by the following partition of unity in the
 subspace of one-particle states:
\be
{\cal P}_1+{\cal P}_2+{\cal P}_{\bot}={\cal I},\quad
{\cal P}_{\bot}={\cal I}-{\cal P}_1-{\cal P}_2,\quad
{\cal P}_i{\cal P}_j=\delta_{ij}{\cal P}_i,
\ee 
\be
{\cal I}=\int u^+({\bf p})\ket{0}\bra{0}u^-({\bf p^{'}})\frac{d{\bf p}}{2p_0},\quad
{\cal P}_{1,2}=\left(\int\psi_{1,2}({\bf p^{'}})u^+({\bf p^{'}})\ket{0}\frac{ d{\bf p^{'}} }{ 2p_{0}^{'} }\right)
\left(\int\bra{0}u^-({\bf p})\psi^{*}_{1,2}({\bf p})\frac{ d{\bf p} }{2p_0}\right)
\ee
For the input state $\ket{\psi_1}$, the probabilities of obtaining
different results are
\be
\mbox{Pr}_1(\psi_1)=\bra{\psi_1}{\cal P}_1\ket{\psi_1}\equiv 1,\quad
\mbox{Pr}_{2,\bot}(\psi_1)=\bra{\psi_1}{\cal P}_{2,\bot}\ket{\psi_1}\equiv 0,
\ee
and similarly for the input state $\ket{\psi_2}$. The measurement defined by
Eqs.(4,5) is non-local in the sense that it requires access to the whole
region of space where the measured field is present. It is intuitively
clear that if we are dealing with the electromagnetic field in an extended
region of space (i.e. the analyzed state is characterized by non-zero
quantum-mechanical averages of the field operators throughout that region
at a certain moment of time), the determination of the field state the
measuring apparatus should be able to probe the field at an arbitrary 
point of the whole region. Even if at any particular point the information 
characterizing the field state gathered due to the local interaction
between the field and the measuring apparatus arises instantly, the transfer
of that information from all the points to a single observer located at a 
certain point of space still requires some time. It is obvious that wherever 
is the observer, this time cannot be less than $L/2c$, where $c$ is the speed 
of light and $L$ is the diameter of the region of non-zero field. Note that
a similar situation also takes place for the systems described by 
non-relativistic quantum mechanics if one takes into account the finite speed
of information transfer. Indeed, consider a composite system consisting
of two (non-interacting) two-level subsystem (particles 1 and 2) located at 
two different points separated by distance $L$. Suppose each of these particles
can be found in one of the two orthogonal basis states known beforehand.
Then, to determine the state of the entire composite system, one should perform
the measurements on both particles. Even if each of these measurements can be 
carried out instantly, the information on their outcomes cannot be conveyed to
a single user, wherever he is located, in time shorter than $L/2c$.

{\it It is only important for the protocol suggested below that in the 
relativistic case two orthogonal states can only be reliably distinguished
in a finite time which depends on their structure. In other words, orthogonal 
states are efficiently indistinguishable (cannot be distinguished reliably)
during a certain finite time interval and become reliably distinguishable
after that time elapses.}

To obtain information on the field state, the measurement should probe the
entire region of space where the field is localized. Therefore, if initially
the field is prepared in a region which is inaccessible for one of the parties
and then propagates to the region accessible for his measuring apparatus,
the state becomes completely accessible only in a finite time. The propagation
amplitude satisfy the causality principle
\be
\BraKet{ \psi_{1,2} (\hat{x}_1) }{ \psi_{1,2} (\hat{x}_2) }=-i
\psi^{*}_{1,2} ( -i\frac{\partial}{\partial {\bf x}_1 } )  
\psi_{1,2}( i \frac{ \partial }{ \partial {\bf x}_2 } )  D^-_0(\hat{x}_1-\hat{x}_2), 
\ee 
where $D^-_0(\hat{x})$ is the negative-frequency function
\be
D_{0}^{-}(\hat{x})=
\frac{i}{(2\pi)^{3/2}}\int d\hat{k}\delta(\hat{k}^2)\theta(-k^0)\exp{(i\hat{k}\hat{x})}=
\frac{1}{4\pi}\varepsilon(x^0)\delta(\lambda),
\ee
\be
\varepsilon(x^{0})=\theta(x^{0})-\theta(-x^{0}),\quad\lambda^2=(x^0)^2-{\bf x}^2;
\ee
here $\ket{\psi_{1,2} (\hat{x})}$ is the state in the ``$\hat{x}$''-representation
\be
\ket{\psi_{1,2} (\hat{x})}=
\int\psi_{1,2}({\bf p})e^{i\hat{p}\hat{x}}u^+({\bf p})\frac{d{\bf p}}{\sqrt{2p_0}}\ket{0}.
\ee
Note that for $\hat{x}_1=\hat{x}_2$ 
\be
\mbox{Pr}_{1,2}(\psi_{1,2})=\bra{\psi_{1,2}}{\cal P}_{1,2}\ket{\psi_{1,2}}=
\left|\BraKet{ \psi_{1,2} (\hat{x}_1) }{ \psi_{1,2} (\hat{x}_2) }|_{\hat{x}_1=\hat{x}_2}\right|^2=
\ee
\bdm
-\left|\psi^{*}_{1,2} ( -i\frac{\partial}{\partial {\bf x}_1 } )  
\psi_{1,2}( i \frac{ \partial }{ \partial {\bf x}_2 } )\right|^2  
\left. D^-_0(\hat{x}_1-\hat{x}_2)D^+_0(\hat{x}_2-\hat{x}_1)\right|_{\hat{x}_1=\hat{x}_2 } 
\edm
In spite of the product of two singular distributions ($D^-_0(\hat{x})=-D^+_0(-\hat{x})$)
occurring at $\hat{x}^2=0$ in Eq.(11), such a product is a correctly defined 
distribution since the convolution of two distributions whose supports lie 
in the front part of the light cone always exists [19].

Let us now describe the protocol. The parties agree beforehand on the states 
$\ket{\psi_1}$ and $\ket{\psi_2}$ corresponding to 0 and 1, respectively. 
The protocol starts at $t=0$ when both parties begin the preparation of 
$N$ states (the number $N$ is also agreed upon beforehand). The worst case
is realized when each user (party), on the one hand, possesses a complete control 
of only the nearest neighbourhood of his own laboratory and, on the other hand,
can deploy his equipment in the immediate vicinity of the laboratory of
the other party in an attempt to cheat him. This means that the protocol
should be stable in the situation when one of the users (parties) can instantly 
convey information to the other user, i.e. when the length of the communication 
channel between them is effectively zero. It is actually sufficient to require
that the efficient size of the region of space where the state is localized
substantially exceeds the communication channel length. Formally, the situation
where the communication channel length is zero is equivalent to the case
where the parties cannot control the space beyond the immediate vicinity of
their laboratories located around the points $x_{A,B}$. At $t=0$ each user
turns on the source of states $\ket{\psi_1}$ and $\ket{\psi_2}$ chosen for
the communication which immediately start to propagate into the communication 
channel and thus become accessible for the measurements. Reliable 
distinguishability (or distinguishability with the probability arbitrarily close
to unit) requires finite time $T$. The reliable distinguishability can be 
achieved employing the measurement described by Eqs.(4,5).

After the time $T/2$ elapses, user A discloses to user B one half ($N/2$)
of the states he has just sent to him. When this information reaches B,
he discloses his $N/2$ states to A, and only after obtaining this 
classical information from B user A discloses the remaining $N/2$ states.
Finally, B discloses his remaining $N/2$ states.

At this stage each user can check consistency between the outcomes of the
quantum-mechanical measurements performed by him and the data publicly 
announced by his counterpart. Each single fault, e.g. when user A announced 
that his $i$-th state was $\ket{\psi_1}$ (0) while user B had his detector 
tuned to $\ket{\psi_2}$ (1) firing means that the protocol is aborted. 
The exchange of classical information and reliable distinguishability of
 orthogonal quantum states make the substitution of even a single bit
impossible which will be important for the subsequent calculation of the 
parity bit. 

Then, if after the exchange of classical information, both parties agree
in the absence of cheating (classical information is fully consistent with
the outcomes of quantum measurements) the parity bit
$c=c_1\oplus c_2\oplus\ldots c_N$ ($c_i=a_i\oplus b_i$, $a_i,b_i$ are the bits
sent by A and B, respectively) is calculated which is
the required random bit the parties A and B wished to generate.

Let us now discuss possible cheating strategies, e.g. for user A.

First of all, exchange of classical information is necessary to exclude the 
possibility of immediately re-sending by user A the states he received from B 
without even trying to analyze them. For the case of photons the latter would 
mean using a mirror mounted by user A just at the point where the 
quantum communication channel used by user B to send his states to user A
leaves his laboratory (i.e. at $x_B$). If the parties had agreed beforehand, 
for example, that the random bit equal to zero means that user A wins, he
could always cheat by simply re-sending the states obtained from B back to him
without even trying to analyze them were it not for the necessity to
disclose the classical information on the states he sent to B later.
Indeed, the parity bit in that case would clearly always be zero (A wins)
since $a_i\equiv b_i$, $c_i=a_i\oplus b_i=b_i\oplus b_i\equiv 0$, 
$c=c_1\oplus c_2\oplus \ldots c_N\equiv 0$. On the other hand, if
user A has to announce through the classical communication channel
which states he actually sent to B, this strategy obviously fails 
because of the no-cloning theorem.

The alternating disclosure of a half of states through a classical communication
channel is necessary to eliminate the following cheating strategy. Since user B
controls only the immediate vicinity of his laboratory (point $x_B$), user A 
can deploy his equipment near $x_B$ and, after re-sending quantum states back to
B, at the stage of exchanging classical information user A can almost instantly 
send back to B the information received from him through the classical channel.
Had user B unveiled all the $N$ states (which user A sent back to him), user A 
would be able to send instantly back to him the classical information received.
The two-stage disclosure of the sent states (one half at a time) allows to 
detect that kind of cheating.

It is important for the protocol that the states are quantum.  If the states 
were classical, the user A could always evade the no-cloning theorem by using 
an infinitesimal fraction of each state sent to him by user B to measure that 
state (which is not prohibited in classical physics) while simultaneously 
sending back these states back to B without disturbing them. The 
collected infinitesimal fraction of the states could be used to 
analyze them during the time $T$ and the information obtained could be sent 
to user B at the stage of exchange of classical information. In that case 
user A always wins. For the quantum states this strategy is impossible
since any measurement disturbs the quantum states.

Since the reliable distinguishability of two orthogonal states
requires a finite time $T$, during the time interval $0\le t\le T$
the states are effectively non-distinguishable (cannot be distinguished
reliably). The probability of the correct state identification,
i.e. the probability of the corresponding detector firing in the time
interval $(0,t)$,
is an increasing function of time $p(t)$ ($p(0)=0$, $p(T)=1$).  
The specific form of function $p(t)$ depends on the particular choice 
of the states and is unimportant in our analysis. The user A cannot 
delay sending of his states since then all should be detected during 
the time $T$ and if they are delayed and $N \gg 1$ there will be detection
events beyond the time interval $0\le t\le T$). One of the possible cheating 
strategies could consist in correcting the states whose transmission
had already been started by user A depending on the outcomes of the
measurements performed over the states received from B. In that case
user A should already have the outcome of the measurement performed over
a state sent by B at his disposal by a certain moment $t$
(which occurs with the probability $p(t)$) while the user B
should have not yet detected the state sent to him by user A
(which occurs with the probability $1-p(t)$).
The probability of successful cheating is therefore
\be 
P_{cheating}=p(t)(1-p(t)).  
\ee 
The maximum of $P$ is reached at $p(t_c)=1/2$ where $P=1/4$ which
is less than the probability of simple guessing which is 1/2.
Therefore, the probability of correct calculation of the parity bit 
encoded in the states sent by B is $P^{N}=(1/4)^N$. Generally,
user A could perform a collective measurement over all $N$ states sent to
him by B simultaneously. Because of the effective non-orthogonality
of the states during the time interval $T$ the probability of success
(see e.g. Ref.[20] on the optimal detection of the parity bit)
in that case is $\sqrt{P^{N}}=(1/2)^N$ which is again not better
than simply guessing at the parity bit.

We conclude with the following remark. The possibility of a reliable 
identification of a state during a finite time interval $T$ depends 
on whether or not there exist the states with a finite spatial support
for the chosen type of particles. In the case of photons only the 
exponentially (with respect to the energy density and detection rate) 
localized states are currently known to exist [21]. The latter formally means 
that the reliable identification (with the probability strictly equal to 1)
can only be achieved with an infinite time interval. However, this consideration 
does not impose any substantial restrictions on the protocol since the time
interval can be chosen sufficiently long to ensure the exponentially close to 
unity probability of the distinguishability of two orthogonal states.

The authors are grateful to A.Kent for useful explanatory
remarks concerning his classical relativistic protocols [15].
This work was supported by the Russian Foundation for Basic Research
(project No 99-02-18127) and by the grant No 02.04.5.2.40.T.50 
within the framework of the Program 
``Advanced devices and technologies in micro- and nanoelectronics''.

\end{document}